# Beyond principlism: Practical strategies for ethical AI use in research practices

**Zhicheng Lin**
Department of Psychology
University of Science and Technology of China

**Correspondence**
Zhicheng Lin, No. 96 Jinzhai Road, Baohe District, Hefei, Anhui, 230026, China
(zhichenglin@gmail.com; X/Twitter: @ZLinPsy)

**Acknowledgments** I thank Yuanqi Du, Sidney K. D'Mello, Benjamin Lira Luttges, Zhan Shi, and Hanchen Wang for their comments on early drafts.
**Competing interests** None declared.
**Funding** The writing was supported by the National Key R&D Program of China STI2030 Major Projects (2021ZD0204200), the National Natural Science Foundation of China (32071045), and the Shenzhen Fundamental Research Program (JCYJ20210324134603010). The funders had no role in the decision to publish or in the preparation of the manuscript.
**AI usage** I used GPT-4o and Claude 3.5 Sonnet for proofreading the manuscript, following the prompts described at https://www.nature.com/articles/s41551-024-01185-8.

**Abstract**
The rapid adoption of generative artificial intelligence (AI) in scientific research, particularly large language models (LLMs), has outpaced the development of ethical guidelines, leading to a "Triple-Too" problem: too many high-level ethical initiatives, too abstract principles lacking contextual and practical relevance, and too much focus on restrictions and risks over benefits and utilities. Existing approaches—principlism (reliance on abstract ethical principles), formalism (rigid application of rules), and technological solutionism (overemphasis on technological fixes)—offer little practical guidance for addressing ethical challenges of AI in scientific research practices. To bridge the gap between abstract principles and day-to-day research practices, a user-centered, realism-inspired approach is proposed here. It outlines five specific goals for ethical AI use: 1) understanding model training and output, including bias mitigation strategies; 2) respecting privacy, confidentiality, and copyright; 3) avoiding plagiarism and policy violations; 4) applying AI beneficially compared to alternatives; and 5) using AI transparently and reproducibly. Each goal is accompanied by actionable strategies and realistic cases of misuse and corrective measures. I argue that ethical AI application requires evaluating its utility against existing alternatives rather than isolated performance metrics. Additionally, I propose documentation guidelines to enhance transparency and reproducibility in AI-assisted research. Moving forward, we need targeted professional development, training programs, and balanced enforcement mechanisms to promote responsible AI use while fostering innovation. By refining these ethical guidelines and adapting them to emerging AI capabilities, we can accelerate scientific progress without compromising research integrity.

## 1 The "Triple-Too" problem of AI ethics

Although scientific workflows and research practices typically change at a glacial pace, this pattern has been disrupted by the recent emergence of generative AI techniques, particularly large language models (LLMs; see **Table 1** for a list of resources that apply generative AI tools to common tasks in scientific research and **Box 1** for a glossary). These transformative techniques are rapidly, if at times invisibly, infiltrating academic corridors, aiding in tasks such as reading, writing, computer programming (coding), idea generation, material creation, data analysis, visualization, and peer review [1-4]. They are also being adopted in scientific, educational, counseling, forensic, and medical applications. This sudden, widespread uptake challenges traditional norms and sparks debates about the role and regulation of generative AI tools in scientific and medical research [5-7].

Yet, a sense of powerlessness grips both ethicists who lament the "uselessness of AI ethics" [8] and users who face ethical uncertainty and confusion, underscoring the necessity and urgency of clear, action-guiding ethical guidelines. As it stands, the current discourse on AI ethics faces what we term the "Triple-Too" problem: *too many* initiatives, frameworks, and documents on high-level ethical directives, which can be overwhelming and difficult to navigate; *too abstract* concepts and principles that lack contextualization and practical relevance, making them hard to implement and follow; and *too negative* a focus on restrictions and risks rather than on benefits and utilities.

To wit, existing AI ethics initiatives primarily focus on establishing general principles for the development, deployment, and governance of AI, addressing developers, consumers, policymakers, and healthcare practitioners [9, 10]. These efforts produce some converging principles such as transparency, justice and fairness, non-maleficence, responsibility, and privacy [10, 11]. Yet, despite numerous efforts to produce ethical frameworks, initiatives, and policies—more than 100 have been identified just between 2015 and 2022 [12]—such abstract AI ethical principles rarely translate into day-to-day practices [13, 14]. Why?

The principle–practice gap exists partly because principles such as fairness are essentially contested and vague concepts—terms that can have conflicting meanings and require context-specific interpretations. For example, fairness could mean equal outcomes across different *demographic* groups (by race, gender, age, and beyond), or it could mean similar outcomes for individuals who are alike in *relevant* aspects (e.g., academic background). In the medical context, for example, fairness criteria such as demographic parity, predictive parity, and equalized odds often conflict with one another in practice [15]. Furthermore, different principles, such as transparency and privacy, can also be conflicting—increasing transparency may involve revealing sensitive data that could violate privacy. Hence, historical lessons from medical ethics teach us that high-level principles must be translated into practices by embedding them into specific norms and cultures. In this regard, AI ethics lacks corresponding norms and cultures to develop best practices for implementing principles and policies.

The predominant focus on abstract principles—often referred to as principlism [16]—is also problematic, given the diversity of AI technologies and sectors involved. Principles suitable for AI governance at the societal level (e.g., socially beneficial AI) might not be pertinent to AI deployment in academic settings (e.g., reproducible AI), and those derived from consumer contexts might be ill-suited for academic contexts [9-11, 13]. Indeed, despite calls for ethical guidelines in academic research [1, 2, 5, 7], discussions have been largely fragmented and without a formative impact on research practices.

Last but not least, current discourse on AI ethics often disproportionately emphasizes restrictions, risks, and harms, sometimes focusing on worst-case scenarios [17]. While there are legitimate concerns—model opacity, potential biases, hallucinations, unreliability, and a host of other limitations [18]—an excessive focus on negativity is counterproductive. It can lead to overly restrictive policies, create negative attitudes and unnecessary fear among potential users, and overshadow the real benefits and opportunities of generative AI in scientific research and education (see **Table 1**). Solving the "Triple-Too" problem requires a shift in perspective and approach.

**Table 1 | Resources for leveraging generative AI tools for common tasks in scientific research**

| Category | Summary | Use case |
|---|---|---|
| Overview [1] | Tools like LLMs are intelligent, versatile, and collaborative—they enhance scientific research by automating and improving text-based tasks (e.g., writing, programming, summarizing, text analysis, generating research ideas) and administrative work (e.g., emails, forms, application materials). | LLMs can generate concise, informative summaries of research papers, helping researchers stay abreast of the latest developments and cite references properly. |
| Prompts [19] | Effective prompting is key to maximizing the utility of LLMs—by crafting precise prompts, adding relevant context, and using examples, users can guide LLMs to produce more accurate, relevant responses. Asking for multiple options, assigning roles, and continuous experimentation further enhance LLM utility. | Instead of asking an LLM to "*Summarize the article*", we might specify: "*Summarize each and every key point/insight, and each must come with rich, supporting details. At the end, explain what is new/novel in the provided text.*" |
| Academic writing [20] | LLMs enhance academic writing through cognitive offloading and imaginative stimulation, offering various levels of assistance—from basic editing to content generation—during idea development (e.g., outlining) and iterative drafting. Specific, tailored prompting maximizes their utility. | An effective copyediting prompt is: "*Offer three revisions. The first is a copyedited version, with the following rules: - only make needed changes; - when the text provided is long, copyedit through each paragraph. The second and third versions need to maximize clarity and flow by following the key recommendations suggested by the following books, but make the two versions distinct: 1) The Elements of* |
| | | *Style; 2) Style: Lessons in Clarity and Grace; 3) On Writing Well; 4) The Sense of Structure: Writing from the Reader's Perspective*." |
| Computer programming (coding) [21, 22] | LLMs transform coding into an intuitive, conversational process, making it more accessible and efficient. By offering six types of assistance—code understanding, generation, debugging, optimization, translation, and learning—they enhance coding practices for both novices and experts. A structured five-step workflow, from project preparation to code refinement, ensures effective AI integration. | For interpreting legacy code or code from others (e.g., published papers), LLMs can offer a high-level overview by highlighting the structure, identifying key algorithms, and explaining how different modules relate to each other. LLMs can further summarize the functionality of various components, provide detailed annotations, and clarify the logic behind specific implementations, making it easier to understand and modify the code. |
| Text analysis (annotation) [23] | Transformer-based models like GPT enable advanced text analysis in social and behavioral sciences, automating tasks such as text classification and inference of psychological variables. | For analyzing social media posts (e.g., a dataset of tweets) to infer psychological variables, language models can classify each tweet into categories of the Big Five personality traits. |

**Box 1 | Glossary of technical AI terms**

**Adversarial Learning**: A training technique where an AI model is trained alongside an adversary model to identify and correct biases or errors, enhancing the robustness of the primary model.

**Adversarial Testing (Red Teaming)**: A method used to expose vulnerabilities in AI systems by challenging them with difficult or edge cases, often to test for biases or weaknesses.

**Catastrophic Forgetting**: A phenomenon where a machine learning model forgets previously learned information when fine-tuned on new tasks or data, leading to decreased performance on earlier tasks.

**Continual Learning**: A method used in machine learning where the model is continually updated or trained on new data without forgetting previously learned tasks (to avoid catastrophic forgetting).

**Customization (Fine-tuning)**: The process of adjusting a pre-trained AI model to perform better on specific tasks or with specific datasets.

**Data Augmentation**: A technique used to artificially expand training datasets by creating modified versions of existing data, such as adding noise, rotating images, or translating text, to improve model robustness.

**Data Debiasing**: Techniques used to reduce biases in training data, such as reweighting or augmenting the data to ensure more equitable representation of different groups.

**Domain Adaptation**: Techniques that adjust a model trained in one domain (data type or environment) to perform well in a different but related domain, addressing distribution shifts or domain-specific biases.

**Embedding**: A method used in machine learning to represent words, phrases, or other types of data as continuous, dense vectors in a high-dimensional space. This allows the model to capture semantic relationships between data points, where similar data points (e.g., words with similar meanings) are located closer together in the embedding space.
**Explainability**: The degree to which an AI model's processes and decisions can be understood by humans. Explainability is crucial for ensuring transparency, trust, and accountability in AI systems, especially in high-stakes applications like healthcare and legal decision-making.
**Federated Learning**: A privacy-preserving technique that allows machine learning models to be trained across decentralized data sources, improving data security by keeping the data local.
**Human in the Loop (HITL)**: An AI system design where humans are actively involved in the decision-making process, providing oversight and the ability to intervene when necessary.
**Human on the Loop (HOTL)**: A system design where humans supervise AI decision-making processes and intervene only when necessary, maintaining oversight but with less active involvement than HITL.
**Human out of the Loop (HOOTL)**: A fully autonomous AI system where human intervention is minimal or nonexistent, posing significant ethical concerns related to accountability and oversight.
**Out-of-Distribution (OOD) Data**: Data that substantially differ from the data used to train an AI model, often leading to poor model performance in real-world applications.
**Retrieval-Augmented Generation (RAG)**: A technique that enhances the accuracy of AI outputs by sourcing relevant information from external databases to inform the model's response.
**Regularization Techniques:** Methods used during model training to prevent overfitting by penalizing overly complex models. This helps the model generalize better and can mitigate learning biases.
**Reinforcement Learning from Human Feedback (RLHF)**: A training technique where human feedback is used to improve an AI system's decision-making. Human evaluators rate or provide input on model outputs, and the model adjusts based on this feedback to better align with human preferences.
**Reweighting**: A bias mitigation technique that assigns higher importance to underrepresented data points during training to improve the model's fairness and accuracy across different groups.
**Style Transfer**: A technique that involves applying transformations to data from different domains to make them more similar, helping models generalize across domains or data distributions.
**Targeted Data Collection**: A technique for deliberately gathering additional data from underrepresented groups or specific domains to improve model performance and reduce bias, ensuring that AI models generalize better to diverse populations.

## 2 A shift to user-centered realism in scientific contexts

While broad, high-level ethical principles such as justice and fairness provide an overarching ethical framework, they can be too abstract to offer actionable guidance, particularly in the complex, rapidly evolving field of generative AI. Similarly, rigid applications of ethical rules or

guidelines—known as formalism [24]—can be too strict to be practical without considering the specific research situation. Resorting to technological fixes alone, by focusing on more advanced models or more sophisticated algorithms—known as technological solutionism (techno-solutionism) [25]—is also unlikely to solve everything, given that AI use is fundamentally rooted in social and cultural contexts. Therefore, we need to move beyond abstract principles by grounding them in practical, context-specific guidelines, adapting them to the specific circumstances at hand, and considering social and cultural contexts. In other words, we need a shift to *realism* [24].

Moving to realism in the context of scientific practices requires us to confront two central, interrelated issues first. One is to clarify the nature of generative AI systems in scientific research: are they best conceptualized as tools or as automatic agents (i.e., systems that can perceive their environment, make decisions, and act to achieve predetermined goals)? A classic definition of AI is "a growing resource of interactive, autonomous, self-learning agency, which enables computational artifacts to perform tasks that otherwise would require human intelligence to be executed successfully" [26]. In other words, AI systems—such as artifacts, models, and technological architectures—possess characteristics commonly associated with agents ("agency") and thus can be agents, but not all AI is inherently agent-like.

Current generative AI tools—such as LLMs (e.g., ChatGPT) and LLM-based code editors (e.g., Cursor)—can help with many specific research tasks (**Table 1**), but they may lack the key characteristics that define an automatic agent (e.g., environment interaction). Nevertheless, "AI scientists" are being developed [27], and in the future, these systems may independently and automatically collect data and test hypotheses (e.g., distributing questionnaires online to poll public opinion), making them full-fledged agents. Such agents pose deep-seated concerns about a possible "new dark age" in which AI agents wield the "authority of the past" (i.e., training data), leading to "paradigm lock-in" and "creative arrest" [17]. However, "dark age" pessimism seems overblown, as AI agents could be designed to challenge past findings and test hypotheses going against the training data. In addition, real-world implementations will likely place humans either *in* the loop (HITL, with active human involvement) or *on* the loop (HOTL, with active human supervision, intervening when necessary), rather than completely *out* of the loop (HOOTL, with minimal or no human intervention). AI scientists ultimately complement rather than replace competent human scientists. Whatever the future holds, we currently have a collection of AI systems with varying degrees of agency at our disposal.

The second issue then is: in the AI ecosystem, what is the ethical role of practicing scientists? AI systems do not meet the criteria for moral agency (e.g., consciousness, free will, moral reasoning, accountability) [28], and responsibility remains with the humans who design, deploy, and use these technologies [29]—a responsibility that is shared in this ecosystem of distributed agency [26]. A practicing scientist may choose to opt out of this ecosystem by not using AI tools, thereby escaping shared responsibility. However, while opting out may seem viable in the short term, it may not be a long-term option and carries two types of ethical and personal risks. Consider the responsibility of scientists as truth seekers and knowledge disseminators: we have much to lose—morally and professionally—if we neglect AI tools that can greatly enhance research quality and productivity. Further, as mentors and educators, we would also miss the opportunity to properly educate students and the next generation of scientists by not engaging with AI tools. Scientists have no moral option to disregard AI tools.

What, then, should a thoughtful scientist do to use AI tools ethically in research practices? Much discussion has focused on the institutional level—establishing principles and rules, governance and oversight, policies and regulations [12-14]. For instance, educational institutions have been implementing various initiatives, including AI detection/watermarking tools, training programs and curricula, guidelines, and AI ethics boards. These efforts are important, as they facilitate norm and culture building [30]. At the same time, if we think of AI as a method of discovery, the pressing issue becomes how to ethically use this method in day-to-day scientific practices. While dealing with risks and harms is an intrinsic part of ethics, an exclusive focus on accountability—from regulation and enforcement to governance control—is hardly effective when benefits and utilities are what motivate most users to adopt new technologies.

Much needed, then, is the development of user-centric ethical guidelines that are balanced—not just focusing on restrictions and risks but also emphasizing benefits and opportunities—and action-guiding, as contextualized in scientific practices. Encouragingly, work in this direction has started to emerge. In particular, Watkins (2023) [31] proposed norms centered on scientific research workflows—from context and embedding models to fine-tuning and agents—with a corresponding checklist for reviewers. Barman et al. (2024) [32] emphasized the importance of providing user education, training, and guidelines on specific LLM use—from writing prompts to evaluating outputs—rather than solely focusing on increasing AI transparency and explainability.

To contribute to this effort, I move beyond principlism and embrace a user-centered, realism-inspired approach. This approach is grounded in five specific goals and actionable strategies for the ethical AI implementation in scientific research and training. These goals and strategies, alongside realistic cases of misuse and corrective measures, are summarized in **Table 2**. Developed by researchers for researchers, they aim to bridge the gap between abstract ethical principles and practical application, facilitating responsible AI use that supports the adoption of generative AI tools and empowers researchers [33].

**Table 2 | Five specific goals, cases of misuse, and action-guiding strategies for ethical AI use in research practices**

| Goal | Case of misuse | Strategy |
|---|---|---|
| Understand model training, fine-tuning, and output | A researcher uses an AI tool for clinical assessments, but the output is not cross-verified with standard diagnostic methods, leading to misclassification of mental health conditions due to unchecked biases in the AI model | Develop a conceptual understanding of how AI models work, including their probabilistic nature, issues related to truthfulness, biases, and inherent opacity. In high-stakes scenarios, prioritize AI tools with explainability features and verify outputs against independent standards or through empirical testing. |
| Respect privacy, confidentiality, and copyright | A research assistant employs an LLM to analyze confidential patient interviews, without ensuring the model's compliance with HIPAA regulations, resulting in a violation of data protection | For sensitive or proprietary information, opt out of model training when possible. If opting out is not an option, consider removing such information in the prompt (e.g., by anonymizing personal data). For fine-tuning with sensitive data, consider using privacy-preserving techniques such as federated learning. When |
| | | incorporating AI-generated content, review and verify the originality of the text. |
| Avoid plagiarism and policy violations | A scientist uses an AI tool for literature review and incorporates AI summaries into a paper without proper citations, raising concerns about plagiarism | Consult and stay up to date with relevant policies (e.g., from journals or institutions). Be transparent, proactive, and reflective: document and disclose AI's role, seek feedback from peers and ethical review boards, and remain vigilant about evolving guidelines, regulations, and policies. When in doubt, err on the side of caution and be prepared to justify ethical decisions made during the research process. |
| Apply AI beneficially as compared with alternatives | A clinical trial uses an AI algorithm to predict patient responses to treatment, but fails to consider the algorithm's limitations in handling diverse patient data, leading to inferior treatment recommendations for certain demographics | Evaluate AI's utility against existing alternatives (or not using AI at all), rather than based on its reliability or performance in isolation. Consider potential benefits, limitations, and negative consequences. In applied research, empirically test applications using objective outcomes or expert/user ratings, rather than assuming benefits and values. |
| Use AI transparently and reproducibly | A team fails to document the specific AI tools and prompts used in their research, making it hard to evaluate or replicate the study | When AI is used beyond basic editing, document key aspects of AI usage in research, including: AI tool and version, specific contributions, data and prompts used, output variability, parameters and settings, model archiving, and iterative documentation. The level of detail in documentation should be balanced with practical considerations and project-specific needs. |

## 3 Five specific goals and action-guiding strategies for ethical AI use in research practices

The goals and strategies presented here derive from my experience as a practicing scientist and AI user and researcher, with two main considerations. On the one hand, I aim to align them with commonly recognized ethical principles (e.g., beneficence, non-maleficence, and justice), grounding them in real-world research challenges to make them context-aware and actionable (e.g., "apply AI beneficially as compared with alternatives"; see Goal four). On the other hand, I avoid being overly prescriptive (e.g., providing checklists) and technical (e.g., breaking down the inner workings of the transformer architecture), to allow the flexibility needed for different research contexts and purposes. While these goals and strategies address key concerns, they are not exhaustive; rather, they offer an initial framework for adaptation and expansion according to the specific research field and the development of generative AI.

I emphasize the importance of resources, such as step-by-step tutorials and new techniques, to help users maximize the potential of generative AI tools [32]; they are, however, often stand-alone contributions—beyond the scope of this discussion. **Table 1** briefly summarizes some of these contributions, which highlight the key benefits that current AI models offer for common research tasks.

### 3.1 Goal one: understand model training, fine-tuning, and output

Responsible and ethical use requires a nuanced understanding of the tool in question. LLMs are adept at generating text in a probabilistic manner—predicting the next token—based on their vast training datasets, often supplemented by reinforcement learning from human feedback (RLHF). Hence, the model output is probabilistic and intricately related to the input, training, customization, and deployment/application, raising major ethical concerns, such as truthfulness and bias. Regarding truthfulness, while methods have been developed to enhance output accuracy, such as retrieval-augmented generation (RAG)—a technique that sources relevant information from a given database to inform the model's response [34]—LLMs can still produce statements that sound confident but are incorrect or misleading (i.e., hallucinations).

Bias is insidious, infiltrating the multistage process of model training [35]. As **Figure 1** and **Box 2** detail, biases can emerge during data collection, preprocessing, model pre-training, customization/fine-tuning, and evaluation stages, with various mitigation strategies and measurement techniques available to address them. The consequences of biases primarily manifest in two forms: representational and allocational harms [36]. *Representational* harms occur when LLMs perpetuate stereotypes or misrepresent certain groups, reinforcing societal prejudices. For example, LLMs may generate text that reinforces stereotypes about gender, race [37], and geographic regions (e.g., associating them with negative attributes like low intelligence or unattractiveness, particularly for regions with lower socioeconomic status [38]).

*Allocational* harms involve the unfair distribution of resources, opportunities, or other outcomes (e.g., education, jobs, loans). This occurs because of distribution shifts—the data distribution in real-world applications can differ substantially from the training data, known as out-of-distribution (OOD, as opposed to in-distribution) data [39]. Particularly when integrating AI within health and behavioral research, cultural sensitivity is crucial to accurately understand and respond to a range of cultural contexts, which may have different interpretations of behavior, language, and mental health. Thus, a lack of data diversity has resulted in AI systems that can be less accurate or even harmful when applied to underrepresented groups. For example, convolutional neural networks (CNNs) trained on radiology images (such as chest X-rays) have been shown to perform worse for underserved populations, such as Hispanic patients and patients on Medicaid, compared with White patients [15].

To address dataset shifts, two general approaches can be used in tandem: targeted data collection and domain adaptation [15]. Targeted data collection involves the deliberate and strategic collection of data from underrepresented groups or populations to improve diversity and balance the representation of different subgroups in the dataset. A more diverse dataset allows the model to learn from a wider range of examples, leading to better generalization and more robust predictions. However, collecting data from underrepresented populations can be resource-intensive and challenging due to privacy concerns, logistical barriers, or the rarity of certain conditions within these populations.

On the other hand, domain adaptation techniques aim to make models robust to variations in data across different domains (e.g., settings with different protocols, populations, or equipment). These techniques include: reweighting (e.g., assigning higher weights to data points from underrepresented or domain-specific subgroups to compensate for their scarcity in the training dataset, thereby helping the model generalize better across domains); adversarial learning (e.g., training the model alongside a competitive model, or adversary, that aims to detect and maximize the loss associated with biases or domain-specific features, thereby forcing the primary model to learn representations that are invariant to domain-specific variations); and style transfer (e.g., applying transformations to data from different domains to make them more similar in appearance or feature distribution, which helps the model learn from a more consistent and unified dataset, even if the original data came from different sources).

In addition to concerns over truthfulness and bias in output, the inner workings of LLMs are largely inscrutable, such as the types of knowledge, reasoning, or goals that the AI uses to generate output. These concerns are exacerbated by the use of closed-source models [33, 40], which obscure fairness and accuracy and potentially lead to ethical lapses, especially in sensitive applications such as mental health and personnel assessments. They also impede the development of transparent AI systems, undermining trust and responsible use. In contrast, making AI processes transparent and comprehensible, as exemplified by interpretable and explainable AI [41, 42], while not a panacea, can bolster trust, facilitate bias mitigation, and promote interdisciplinary collaboration. The incorporation of AI into individual applications would also benefit from personalized measures of uncertainty [43]. This approach ensures that AI tools are ethically, culturally, and personally attuned.

The key takeaway is that, when integrating generative AI such as LLMs into scientific practices, users must develop a conceptual understanding of how these models work—their probabilistic nature, issues related to truthfulness, the array of biases (sources, types, measures, mitigation strategies, and potential consequences), and their inherent opacity (e.g., challenges with interpretability and explainability, differences between open-source and closed-source models). In high-stakes scenarios, as highlighted in **Table 2**, practitioners should prioritize AI tools with explainability features and, crucially, verify outputs against independent standards or through empirical testing. While generative AI has its limitations, on balance these concerns generally do not pose major risks in routine practices, as outlined in **Table 1**.

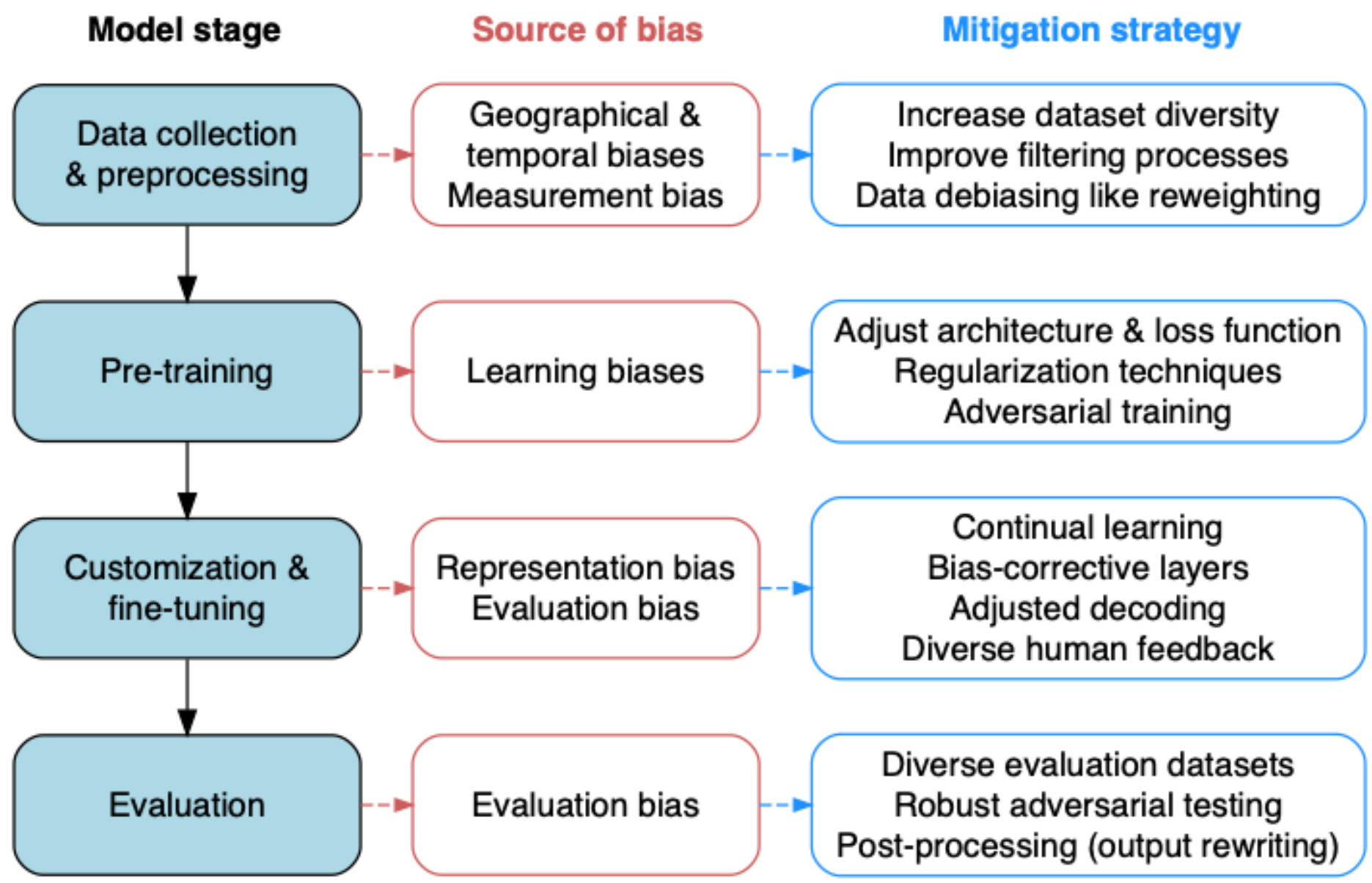

**Figure 1 | Bias introduction and mitigation in large language models.** Biases infiltrate at various stages, starting with data collection and preprocessing, where geographical, temporal, and measurement biases can emerge. During pre-training, learning biases may be ingrained in the model's architecture. Customization and fine-tuning, including RLHF, introduce representation biases, while evaluation biases may arise during both model evaluation and deployment. Mitigation strategies are depicted alongside each stage, such as increasing dataset diversity, applying regularization techniques, ensuring diverse human feedback, and implementing robust adversarial testing. For details, see **Box 2**.

**Box 2 | Bias in large language models (LLMs): Sources, mitigation, and measurement**

As **Figure 1** shows [35, 36], bias first enters during data scraping and sampling, where the selection of online texts—often laden with historical and present-day prejudices—can skew representations across languages, eras, and authors, introducing *geographical* and *temporal biases*. Bias further arises during preprocessing, as methods like decontamination and filtering attempt to cleanse datasets of toxic language; if these tools are flawed or biased, they implant *measurement biases*. Bias mitigation at this stage involves strategies like increasing the diversity of the dataset through data augmentation, improving filtering processes to remove harmful content, and using data debiasing techniques such as reweighting (e.g., adjusting the weight of underrepresented groups) to correct imbalances and ensure fairer representation across different groups.

In the pre-training phase, optimizing the model's weights to minimize loss functions may unintentionally magnify these biases, ingraining them within the model's architecture and manifesting as *learning biases* in word and contextual embeddings. Addressing these issues involves adjustments to the model's architecture or loss function to reduce the impact of these biases, ensuring the model learns in a more balanced way. Regularization techniques, such as adding penalties to the model's loss function to prevent overly complex models, and adversarial training, which involves exposing the model to challenging examples that might trigger biased responses, can further help mitigate biases by making the model less sensitive to biased patterns in the data.

As LLMs transition to customization and deployment, additional biases emerge. Fine-tuning, which tailors the LLM for specific applications using domain-specific datasets, may embed *representation biases* from the datasets, and risk catastrophic forgetting, whereby the model sheds general knowledge, thereby compounding biases. Mitigation strategies during fine-tuning include continual learning approaches to prevent catastrophic forgetting, as well as using bias-corrective layers or regularization techniques to maintain fairness. During text generation, when the model converts embeddings into text, decoding strategies can be adjusted to further reduce the propagation of biases. The RLHF phase introduces *evaluation* biases from human raters, whose perspectives can skew the model's outputs. Fine-tuning in response to such feedback may also exacerbate *representation, measurement, and learning biases*. Mitigation requires diverse and representative human feedback, along with careful monitoring and iterative refinement.

Lastly, the evaluation stage risks *evaluation bias* if benchmark datasets are not representative of the diverse real-world scenarios the model will encounter or if adversarial testing (red teaming)—where the model is challenged with difficult or edge cases to expose weaknesses—is not robust, allowing biases to enter or remain undetected. To help detect and correct biases before the model is fully deployed, it's crucial to use diverse and comprehensive evaluation datasets, implement robust adversarial testing, and apply post-processing techniques such as output rewriting (where biased outputs are identified and revised) or calibration adjustments (where the model's predictions are adjusted to reduce bias).

How to measure biases in LLMs? Biases can be measured at multiple levels: internal representations (embedding-based), decision-making probabilities (probability-based), and final outputs (generated text-based). *Embedding*-based metrics assess biases in the representation space by examining the relationships between words and concepts within the model's embeddings (e.g., how certain groups are stereotypically associated with particular attributes). *Probability*-based metrics evaluate biases in the model's assigned probabilities, measuring how likely a model is to produce biased outputs based on different inputs. Generated *text*-based metrics focus on the biases present in the actual outputs (e.g., the frequency and context of biased language). These measures help identify and quantify biases, which is useful for developing targeted strategies for mitigation.

**3.2 Goal two: respect privacy, confidentiality, and copyright**

Interacting with LLMs raises several privacy, confidentiality, and copyright concerns related to both input and output. Input concerns arise when users enter sensitive or proprietary information into their prompts. Unless users opt out, these prompts may become part of the training data, a fact unknown to many users [7]. For example, in ChatGPT, the policy is that by default, "*[w]hen you use our services for individuals such as ChatGPT or DALL•E, we may use your content to train our models*" (https://tinyurl.com/3jry5t4h). In Claude, user inputs are used for model training when the conversations are flagged for review or when users report the materials to Claude (https://tinyurl.com/mrxn62pw).

Challenges arise from two sources, depending on the origin of the input information: data from others and data from oneself. In cases involving others' data, when the input contains personally identifiable information (PII) or confidential data from patients or clients, there could be a breach of privacy or confidentiality. This issue is particularly concerning in fields like biomedical and professional services (such as finance and legal work), where compliance with regulations like HIPAA is essential [44]. Additionally, if the input contains copyrighted or otherwise confidential materials—such as uploading content from a published paper or a manuscript under review—this could lead to potential copyright infringements. Although the legality of training on copyrighted materials is still undecided (see the ongoing lawsuit between The New York Times and OpenAI), it is prudent to be proactive. Similarly, users may inadvertently relinquish certain privacy rights over their own content, such as drafts of manuscripts and research data.

On the output side, ethical and legal concerns also arise surrounding the copyright ownership of AI-generated content. Though rare, there have been instances in which the output from LLMs closely resembles existing texts, such as articles from The New York Times or other copyrighted materials. Major AI providers like OpenAI, the provider of ChatGPT, transfer ownership of the model output to the user; similarly, Microsoft and Google, providers of Bing and Gemini, respectively, offer protection against legal risks. Nevertheless, the debate and legal status regarding the copyright of AI-generated content remains unsettled.

What are the potential solutions? For sensitive or proprietary information, one strategy is to opt out of model training when possible (for ChatGPT, see the link above or submit privacy requests here: https://privacy.openai.com/policies). If opting out is not an option, consider removing such information from the prompt (e.g., by anonymizing personal data). For fine-tuning with sensitive data, consider using privacy-preserving techniques such as federated learning [15], which allows models to train on decentralized data without direct access. When incorporating AI-generated content into our work, we must be diligent in reviewing and verifying the originality of the text, including checking it against existing literature.

### 3.3 Goal three: avoid plagiarism and policy violations

Aside from the legal aspects of AI use—privacy, confidentiality, and copyright—ethical concerns also arise when using model output in one's work. These concerns include potential plagiarism and policy violations. While most journals permit AI usage, the extent of allowable use varies [7]. For example, *Nature*, *Science*, and their family journals ban AI-generated images in manuscripts; journals from Elsevier and the *Lancet* group allow AI solely to enhance readability and language in manuscripts [7]. Thus, it is crucial to stay informed and proactive regarding the AI policies of target journals when using AI in manuscript preparation.

In addition to policy considerations, using AI output raises questions of attribution and plagiarism [20]. Without proper attribution or verification of originality, there is a risk of inadvertently committing plagiarism. But how should we attribute the contributions of AI? While AI can contribute substantially to research, the consensus is against granting it authorship [45]. This stance is primarily due to AI's lack of agency and accountability: it cannot be held responsible for errors or ethical considerations. This is a key distinction separating it from posthumous authors, who could have assumed such responsibilities during their lifetimes.

Properly attributing AI contributions, however, is not always straightforward. For instance, consider a user incorporating an AI-generated paragraph into a manuscript. Does this constitute plagiarism? Traditional definitions of plagiarism involve appropriating another *person*'s intellectual work as one's own. Citing AI in references is problematic, and plagiarism may not directly apply because AI is not human, and its output stems from human prompts, with the final product often being a collaboration between the AI and the user. Yet, without proper attribution, it could misrepresent the originality of the work. This concern is not limited to writing but also applies to using novel ideas generated through AI brainstorming.

On the other hand, it is important to recognize that there are no absolute criteria for distinguishing between acceptable use and potential misappropriation of AI contributions. The level of AI contribution may depend on various factors, including the originality of the prompt, the extent of editing or modification of the AI-generated content, and its overall contribution to the final work. These factors are inherently subjective. For example, author Rie Kudan recently received a prestigious Japanese literary award, the Akutagawa Prize, for her sci-fi novel, of which 5% was written by ChatGPT (https://www.vice.com/en/article/k7z58y/rie-kudan-akutagawa-prize-used-chatgpt). Whether it is ethical to include a proportion of a manuscript written by AI in academic publishing remains debatable. Such ambiguities highlight ethics as a process rather than a destination—serious debates are needed, and intractable disagreements are to be expected [13].

How, then, should one navigate this evolving ethical landscape? While it is clear that AI cannot be granted authorship and that AI-generated content should not be misrepresented as original work, other aspects remain murky, particularly the appropriate level of AI involvement and its attribution. To handle this ambiguity, researchers should first and foremost consult and stay up-to-day with relevant policies (e.g., from journals or institutions). Be transparent, proactive, and reflective: document and disclose AI's role, seek feedback from peers and ethical review boards, and remain vigilant about evolving guidelines, regulations, and policies. When in doubt, err on the side of caution and be prepared to justify the ethical decisions made during the research process.

### 3.4 Goal four: apply AI beneficially as compared with alternatives

Given their limitations as well as potential legal and ethical concerns, we may ask: should we use AI tools at all? Or, more specifically, how should we evaluate the utility of AI? An intuitive reaction is to look at the absolute performance—rejecting AI use when the performance is not perfect or reliable. However, a more valid approach would be to compare it with the best available alternative (or with not using AI at all), considering their potential benefits, limitations, and negative consequences of each. When used in research processes, such as writing, coding, and data analysis, AI serves as an additional tool. The benefits include improved efficiency and productivity [20]. Limitations involve understanding the nuances of model behaviors (e.g., hallucinations, bias), learning to prompt effectively (prompt engineering) [19], and navigating the ethical landscape. Potential negative consequences might include overreliance on AI and repercussions from its inappropriate use [20].

In other applications, including in substantive tasks and consequential decision-making, nuanced considerations of costs and benefits are needed [46]. For example, in social sciences, text analysis is a common yet time-consuming and error-prone task. Although AI may not be completely reliable (i.e., its responses do not remain the same for the same prompt) and may not have high accuracy, it can be ideal for the job compared with alternatives, such as crowd workers or trained research assistants—which can be time-consuming, expensive, and not necessarily more reliable or accurate [23]. Similarly, in applied research that evaluates interventions, such as in healthcare or education, the guiding principle should be whether the application of AI is superior to current alternatives (though not necessarily the best possible option). This is particularly relevant in resource-limited regions: AI might be inferior to the best therapists or educators but still be the superior choice, considering availability and its unique strengths—low deployment cost, being patient, non-judgmental, and widely available.

At the same time, enthusiasm must be tempered with careful consideration of potential negative consequences, including safety, efficacy, and fairness. In healthcare, it is vital to "do no harm"—interventions should be evidence-based, as with other interventions [47]. Therefore, applications must be empirically tested using objective outcomes or expert/user ratings, rather than assuming their benefits and values. For example, as a psychotherapist, how accurate is the AI at inferring the emotional state of the user (e.g., from written text)? How should the AI respond to user input? Are the responses beneficial to the user in both the short and long term?

Likewise, when AI is used to make important decisions, such as in personnel evaluation and student admissions or assessments, fairness is crucial. Acknowledging its definitional plurality [48], it is essential to verify that these systems do not inadvertently favor or disadvantage any group due to biased data or algorithms [49]. In practice, AI systems might deliver differential benefits for different groups, so equal outcomes may not be possible, and selective deployments might be more practical [50]. In student assessments, AI needs to be calibrated to recognize and value diverse linguistic expressions and cultural perspectives. In personnel evaluations, algorithms should be scrutinized to avoid biases associated with personal attributes (e.g., gender, ethnicity, or age).

The takeaway message is that, despite its limitations, the ethical application of AI in both basic and applied research entails evaluating its utility against existing alternatives (or not using AI at all), rather than basing the evaluation on its reliability or performance in isolation. This evaluation strategy is in line with the ethical principles of beneficence and non-maleficence.

### 3.5 Goal five: use AI transparently and reproducibly

The dynamic and evolving nature of LLMs presents a unique challenge in ensuring transparency and reproducibility [7]. Iterations of the same model may produce differing outcomes. This variability underscores the necessity for clear documentation of the specific model version and date. Such record-keeping enables other researchers to understand the context and potential limitations of the findings and, where feasible, to replicate the study with the same or similar AI toolsets.

Adding another layer of complexity to reproducibility is the stochastic nature of AI algorithms. Their inherent randomness means that, even with the same input, the output can vary. This unpredictability, while beneficial in generating diverse and creative responses, poses a challenge for producing consistent and repeatable results. Therefore, it is helpful to document not only the AI model and its version but also the specific prompts used, the range of responses received, and the selection criteria for the final output. This documentation will enhance the reproducibility of AI-assisted research, allowing others to understand the decision-making process involved in selecting AI-generated content.

However, this quest for transparency and reproducibility must be balanced with practical considerations, particularly in terms of feasibility and workload. The iterative and extensive nature of AI involvement makes complete documentation a formidable task. Algorithms may undergo multiple adjustments and training phases during the research process, making it difficult to capture every step in a readily replicable manner.

In the context of publishing, the white paper released on December 5, 2023, by the International Association of Scientific, Technical, and Medical Publishers (STM), states that disclosure is not necessary when "[u]ing publicly available GenAI as a basic tool that supports authors in refining, correcting, formatting, and editing texts and documents", but "[a]uthors must disclose any use of GenAI that transcends those use cases so an editorial decision can be made as to its legitimacy" (https://www.stm-assoc.org/new-white-paper-launch-generative-ai-in-scholarly-communications/).

This policy seems sensible, and a consensus is needed to develop field-specific guidelines and implementations [7]. However, the policy does not specify how to properly disclose AI use beyond basic editing, deferring this to journal policies, which often lack clarity and consistency [7]. To complement the STM recommendations and facilitate consensus building, **Table 3** outlines key aspects of AI usage in research, from basic tool identification to detailed documentation of model iterations.

**Table 3 | Documenting AI involvement in research**

| Item | Description | Rationale | Example |
|---|---|---|---|
| AI tool and version | Specify the exact version of the AI tool used | Different versions of the same tool can vary in their capability and behavior | OpenAI GPT-4-Turbo, Feb. 2024 |
| Specific contributions | Indicate which parts were generated or influenced by AI | Specifying the role of AI in the work improves transparency | Abstract, Fig 2, Discussion section were edited by AI |
| Data and prompts | Disclose the training data and any prompts used | Data and prompts strongly affect AI output | "Copyedit the text" [specific passage from paper] |
| Output variability | Report output variations generated across iterations | Unlike traditional software, AI output has built-in randomness | Three sample paragraphs with differing analogies were generated |
| Parameters and settings | Disclose key parameters, settings, and configurations. | Different parameters and setting lead to different results | Temperature 0.85 |
| Archiving models | Make trained models publicly available in repositories | Making the model available facilitates scrutiny and cumulative exchanges | Custom classifier model was uploaded to HuggingFace Hub |
| Iterative documentation | Keep a record of AI usage over multiple iterations | Model performance may depend on the iterative search for | Training execution log detailing parameter |

| Item | Description | Rationale | Example |
|---|---|---|---|
| | | optimal prompts and training | adjustments over 10 iterations |

*Note.* Not all items are necessary to document, and the importance of each item depends on the specific project.

## 4 Concluding remarks

Generative AI has proven quite useful for various research tasks (**Table 1**), but the lack of consensus on AI ethics in research practices leads to a laissez-faire approach to its use, risking misuses (**Table 2**). Indeed, the current discourse on AI ethics falls into the "Triple-Too" trap: *too many* documents on ethical principles; concepts and principles that are *too abstract*; and *too negative* a focus on restrictions and risks. This is perhaps expected, given that the use of generative AI in academic research is only recent [18], without a long professional history and established norms to draw from. Translating abstract principles into practice requires norm and culture building [13], going beyond principlism, formalism, and technological solutionism. Building on previous arguments about the importance of checklists [31] and user education [32, 33] for specific LLM use, I have embraced a user-centered, realism-inspired approach to provide a practical framework. As outlined in **Table 2**, this approach formulates five specific goals, offering action-guiding strategies and analyses of scientific research practices.

Looking ahead, two critical areas emerge for immediate action: professional development and training, and the enforcement of ethical standards. Training should cover both the operational aspects of AI tools and the ethical implications of AI systems—both benefits and risks. However, the establishment of guidelines and training programs is effective only if accompanied by reasonable enforcement mechanisms [1]. Regulatory bodies should develop and implement clear policies regarding AI misuse. This could include periodic audits of AI usage, transparent reporting requirements for AI-assisted work, and the establishment of an ethics review board specializing in AI applications. Enforcement mechanisms should strike a balance between promoting innovation and safeguarding ethical standards.

By focusing on continually refining these ethical guidelines, adapting training programs to emerging AI advancements, and ensuring that enforcement mechanisms are both reasonable and effective, the emergent capabilities of AI could markedly accelerate human understanding without eroding the integrity of science.

**References**

1. Lin, Z., *Why and how to embrace AI such as ChatGPT in your academic life.* Royal Society Open Science, 2023. **10**: p. 230658.
2. Wang, H., et al., *Scientific discovery in the age of artificial intelligence.* Nature, 2023. **620**(7972): p. 47-60.
3. Liang, W., et al., *Can large language models provide useful feedback on research papers? A large-scale empirical analysis.* NEJM AI, 2024. **1**(8).
4. Fecher, B., et al., *Friend or foe? Exploring the implications of large language models on the science system.* AI & Society, 2023.
5. Birhane, A., et al., *Science in the age of large language models.* Nature Reviews Physics, 2023. **5**: p. 277-280.
6. Jecker, N.S., et al., *AI and the falling sky: Interrogating X-Risk.* Journal of Medical Ethics, 2024.
7. Lin, Z., *Towards an AI policy framework in scholarly publishing.* Trends in Cognitive Sciences, 2024. **82**(2): p. 85-88.
8. Munn, L., *The uselessness of AI ethics.* AI and Ethics, 2023. **3**(3): p. 869-877.
9. Badal, K., C.M. Lee, and L.J. Esserman, *Guiding principles for the responsible development of artificial intelligence tools for healthcare.* Communications Medicine, 2023. **3**(1): p. 47.
10. Correa, N.K., et al., *Worldwide AI ethics: A review of 200 guidelines and recommendations for AI governance.* Patterns, 2023. **4**(10): p. 100857.
11. Jobin, A., M. Ienca, and E. Vayena, *The global landscape of AI ethics guidelines.* Nature Machine Intelligence, 2019. **1**(9): p. 389-399.
12. Ghoz, L. and M. Hendawy, *An Inventory of AI ethics: Analyzing 100 documents.* MSA Engineering Journal, 2023. **2**(2): p. 647-675.
13. Mittelstadt, B., *Principles alone cannot guarantee ethical AI.* Nature Machine Intelligence, 2019. **1**(11): p. 501-507.
14. Rees, C. and B. Muller, *All that glitters is not gold: Trustworthy and ethical AI principles.* AI and Ethics, 2022: p. 1-14.
15. Chen, R.J., et al., *Algorithmic fairness in artificial intelligence for medicine and healthcare.* Nature Biomedical Engineering, 2023. **7**(6): p. 719-742.
16. Prem, E., *From ethical AI frameworks to tools: A review of approaches.* AI and Ethics, 2023. **3**(3): p. 699-716.
17. Leslie, D., *Does the sun rise for ChatGPT? Scientific discovery in the age of generative AI.* AI and Ethics, 2023.
18. Resnik, D.B. and M. Hosseini, *The ethics of using artificial intelligence in scientific research: New guidance needed for a new tool.* AI and Ethics, 2024.
19. Lin, Z., *How to write effective prompts for large language models.* Nature Human Behaviour, 2024. **8**(4): p. 611-615.

20. Lin, Z., *Techniques for supercharging academic writing with generative AI.* Nature Biomedical Engineering, 2024.
21. Merow, C., et al., *AI chatbots can boost scientific coding.* Nature Ecology & Evolution, 2023. **7**(7): p. 960-962.
22. Perkel, J.M., *Six tips for better coding with ChatGPT.* Nature, 2023. **618**(7964): p. 422-423.
23. Rathje, S., et al., *GPT is an effective tool for multilingual psychological text analysis.* Proceedings of the National Academy of Sciences of the United States of America, 2024. **121**(34): p. e2308950121.
24. Green, B. and S. Viljoen, *Algorithmic realism: Expanding the boundaries of algorithmic thought*, in *Proceedings of the 2020 Conference on Fairness, Accountability, and Transparency*. 2020, Association for Computing Machinery: Barcelona, Spain. p. 19–31.
25. Pham, B.-C. and S.R. Davies, *What problems is the AI act solving? Technological solutionism, fundamental rights, and trustworthiness in European AI policy.* Critical Policy Studies, 2024: p. 1-19.
26. Taddeo, M. and L. Floridi, *How AI can be a force for good.* Science, 2018. **361**(6404): p. 751-752.
27. Lu, C., et al., *The AI scientist: Towards fully automated open-ended scientific discovery.* arXiv:2408.06292, 2024.
28. Coeckelbergh, M., *Artificial Intelligence, Responsibility Attribution, and a Relational Justification of Explainability.* Science and Engineering Ethics, 2020. **26**(4): p. 2051-2068.
29. Matthias, A., *The responsibility gap: Ascribing responsibility for the actions of learning automata.* Ethics and Information Technology, 2004. **6**(3): p. 175-183.
30. Seger, E., *In defence of principlism in AI ethics and governance.* Philosophy & Technology, 2022. **35**(2): p. 45.
31. Watkins, R., *Guidance for researchers and peer-reviewers on the ethical use of Large Language Models (LLMs) in scientific research workflows.* AI and Ethics, 2023.
32. Barman, K.G., N. Wood, and P. Pawlowski, *Beyond transparency and explainability: On the need for adequate and contextualized user guidelines for LLM use.* Ethics and Information Technology, 2024. **26**(3): p. 47.
33. Lin, Z., *Progress and challenges in the symbiosis of AI with science and medicine.* European Journal of Clinical Investigation, 2024. **n/a**(n/a): p. e14222.
34. Lewis, P., et al., *Retrieval-augmented generation for knowledge-intensive NLP tasks*, in *Advances in Neural Information Processing Systems*. 2020. p. 9459-9474.
35. Lee, J., et al., *The life cycle of large language models in education: A framework for understanding sources of bias.* British Journal of Educational Technology, 2024. **55**(5): p. 1982-2002.

36. Gallegos, I.O., et al., *Bias and fairness in large language models: A survey.* Computational Linguistics, 2024: p. 1-83.
37. Fang, X., et al., *Bias of AI-generated content: an examination of news produced by large language models.* Scientific Reports, 2024. **14**(1): p. 5224.
38. Manvi, R., et al., *Large language models are geographically biased.* arXiv:2402.02680, 2024.
39. Chen, Z., W. Li, and Z. Lin, *Characterizing patients who may benefit from mature medical AI models.* PsyArXiv preprint, 2024.
40. Palmer, A., N.A. Smith, and A. Spirling, *Using proprietary language models in academic research requires explicit justification.* Nature Computational Science, 2023.
41. Garrett, B.L. and C. Rudin, *Interpretable algorithmic forensics.* Proceedings of the National Academy of Sciences of the United States of America, 2023. **120**(41): p. e2301842120.
42. Ehsan, U. and M.O. Riedl, *Social construction of XAI: Do we need one definition to rule them all?* Patterns, 2024. **5**(2): p. 100926.
43. Banerji, C.R.S., et al., *Clinical AI tools must convey predictive uncertainty for each individual patient.* Nature Medicine, 2023. **29**(12): p. 2996-2998.
44. Ong, J.C.L., et al., *Medical ethics of large language models in medicine.* NEJM AI, 2024. **1**(7): p. AIra2400038.
45. Stokel-Walker, C., *ChatGPT listed as author on research papers: many scientists disapprove.* Nature, 2023. **613**(7945): p. 620-621.
46. Ugar, E.T. and N. Malele, *Designing AI for mental health diagnosis: challenges from sub-Saharan African value-laden judgements on mental health disorders.* Journal of Medical Ethics, 2024.
47. Sharma, A., et al., *Human–AI collaboration enables more empathic conversations in text-based peer-to-peer mental health support.* Nature Machine Intelligence, 2023. **5**(1): p. 46-57.
48. Verma, S. and J. Rubin, *Fairness definitions explained*, in *Proceedings of the International Workshop on Software Fairness*. 2018, Association for Computing Machinery: Gothenburg, Sweden. p. 1–7.
49. Lira, B., et al., *Using artificial intelligence to assess personal qualities in college admissions.* Science Advances, 2023. **9**(41): p. eadg9405.
50. Goetz, L., et al., *Generalization—a key challenge for responsible AI in patient-facing clinical applications.* npj Digital Medicine, 2024. **7**(1): p. 126.